\documentclass[aps,superscriptaddress,showpacs,nofootinbib,eqsecnum,prd,notitlepage,twocolumn]{revtex4-1} 

\usepackage{graphicx}
\usepackage{bm}
\usepackage{dcolumn}
\usepackage{amssymb}
\usepackage{amsmath}
\usepackage{epsfig}
\usepackage[english]{babel}
\usepackage[latin1]{inputenc}
\usepackage{eucal}
\usepackage{hyperref}
\usepackage{verbatim}
\usepackage{latexsym}

\begin{document}

\title{Thermodynamics of quark matter with a chiral imbalance}

\author{Ricardo L. S. Farias} 
\affiliation{Departamento de F\'{\i}sica, Universidade Federal de Santa Maria, 
97105-900, Santa Maria, RS, Brazil}
\affiliation{Physics Department, Kent State University, Kent, OH 44242, USA} 

\author{Dyana C. Duarte}
\affiliation{Departamento de F\'{\i}sica, Universidade Federal de Santa Maria, 
97105-900, Santa Maria, RS, Brazil}

\author{Gast\~ao  Krein} 
\affiliation{Instituto de F\'{\i}sica Te\'orica, Universidade Estadual
  Paulista,  01140-070 S\~ao Paulo, SP, Brazil}

\author{Rudnei O.  Ramos}
\affiliation{Departamento de F\'{\i}sica Te\'orica, Universidade do
  Estado do Rio de Janeiro, 20550-013 Rio de Janeiro, RJ, Brazil}  

\begin{abstract}

We show how a scheme of rewriting a divergent momentum integral
can conciliate results obtained with the Nambu--Jona-Lasinio model and
recent lattice results for the chiral transition in the presence of a
chiral imbalance in quark matter. Purely vacuum contributions are separated 
from medium-dependent regularized momentum integrals in such a way that one is 
left with ultraviolet divergent momentum integrals that depend on vacuum 
quantities only. The scheme is applicable to other commonly used effective 
models to study quark matter with a chiral imbalance, it allows us to identify 
the source of their difficulties in reproducing the qualitative features
of lattice results, and enhances their predictability and uses in other 
applications. 
\end{abstract}


\maketitle

\section{Introduction}

There has been an increased interest recently in the study of how a
chiral imbalance of right-handed and left-handed quarks can influence
the phase diagram  of quantum chromodynamics (QCD). There are many
good reasons for  this interest. {}For instance, the nontrivial nature
of the vacuum  of non-Abelian gauge theories in general, and of QCD in
particular,  allows for the existence of topological solutions like
instantons and  sphalerons. While instantons describe the quantum
tunneling between different  vacua, sphalerons are classical solutions
describing transitions going above the  barrier between the
vacua. Sphaleron processes are unsuppressed at high
temperatures~\cite{Gross:1980br,Aoyama:1987nd} and, from the
Adler-Bell-Jackiw anomaly, they can generate, in the context of QCD, an
asymmetry between the number of left- and right-handed quarks. Such a
chirality imbalance is  expected to occur in  event-by-event $C-$ and
$CP-$violating processes in heavy-ion
collisions~\cite{Kharzeev:2004ey,Fukushima:2008xe}. Moreover, in
off-central collisions  a magnetic field is created and the presence
of a chiral imbalance gives rise to an  electric current along the
magnetic field, whose effect is to produce a charge separation,  an
effect dubbed chiral magnetic effect (CME) in the
literature{\textemdash} see, e.g.,
Refs.~\cite{cme1,cme3,Kharzeev:2015kna,Kharzeev:2015znc} for recent
reviews and references therein.   The CME effect is not restricted to
QCD, it extends over a wide range of systems, e.g., hydrodynamics
and condensed matter
systems~\cite{son2,metlitski,yee,chernodub13,chernodub14,chernodub15,chernodub16},
and it  has been actually observed in many recent condensed matter
experiments~\cite{Li:2014bha},  which makes it of much wider interest
in physics.

The effects of a chiral imbalance in the phase diagram of QCD can be
studied in the  grand canonical ensemble by introducing a
chiral chemical potential $\mu_5$ through  a term $\mu_5{\bar \psi}
\gamma_0 \gamma_5 \psi$ in the QCD Lagrangian
density~\cite{Fukushima:2008xe}. Besides of the intrinsic motivation
in the context of the physics of heavy-ion collisions, there have been
interesting  suggestions~\cite{Ruggieri:2011xc,Gatto:2011wc} that the
phase diagram of QCD in the  $T-\mu_5$ plane could be in principle
mapped into the real phase diagram in the $T-\mu$  plane, where $\mu$
is the usual quark baryon chemical potential, a feature that would
help to pinpoint the expected critical end point (CEP) of QCD
{\textemdash} see Refs.~\cite{Wang:2015tia,Xu:2015vna} for opposite
views. More important, however, is the fact that QCD in the presence
of a chiral chemical potential is free from the  sign problem and,
therefore, amenable to Monte Carlo sampling in lattice simulations,
contrary to the case of QCD in the presence of a baryon chemical
potential, which has the sign problem. Hence, there is hope that
lattice simulations of QCD with $\mu_5$ can be  used as a possible
benchmark platform  for comparing different effective models used in
the literature. In this respect, it is intriguing that models that
have been very successful in describing many features predicted by
universality arguments and lattice simulations for the chiral
transition in QCD at  nonzero $T$ and $\mu$, have difficulties in
reproducing, even at a qualitative level, recent  lattice
results~\cite{Braguta:2015owi,Braguta:2015zta} for the chiral critical
transition temperature  $T_c$ at finite~$\mu_5$. {}For instance,
predictions based on Nambu--Jona-Lasinio (NJL)-type  of
models~\cite{fukushimaPNJL,huang13,Ruggieri:2011xc,huang14,huang15,craigmu5pnjl}
and quark linear sigma  models~\cite{chernodub,Ruggieri:2011xc} find
that $T_c$ decreases with~$\mu_5$, while the  lattice results of
Refs.~\cite{Braguta:2015owi,Braguta:2015zta} find $T_c$ increasing
with~$\mu_5$.  

A nonzero quark condensate mixes right- and left-handed
quarks and has the effect of decreasing  the chiral
asymmetry. Therefore, as one forces a system to increase the
right-left asymmetry by increasing $\mu_5$, one expects that the
quark condensate will increase and, therefore, $T_c$  is expected to
increase likewise. This is because addition of left- and right-handed 
quarks to a system, in amounts controlled by $\mu_5$, favors quark-antiquark
pairing, that is, increases the quark condensate~\cite{kotovmu5cat}. 
Universality arguments in the large $N_c$ limit 
(where $N_c$ is the number of color degrees of
freedom)~\cite{yamamotonc} also predict a $T_c$ increasing
with~$\mu_5$. Some recent studies using phenomenological quark-gluon
interactions in the framework  of the Dyson-Schwinger equations for
the quark propagator~\cite{Wang:2015tia,Xu:2015vna} and nonlocal
finite-range NJL  models~\cite{ruggierimu52,frasca} find a $T_c$
increasing with $\mu_5$. Both types of models  have in common the
feature of having a momentum-dependent quark mass function, in
contrast to  a constant mass in contact-interaction models. A
qualitative  agreement with the lattice results for $T_c$ was also
found in Ref.~\cite{ruggierimu51}, by using a nonstandard
renormalization scheme in the quark linear sigma model. 

Given the prominent role played by NJL type of models in providing
insight into the problem of the chiral phase transition, it is
important to identify the sources of their failure  in reproducing the
qualitative features of lattice simulations for the $\mu_5$ dependence
of $T_c$.  In the present work we pursue such a study. Our analysis is
based on a proper separation  of medium effects from divergent
integrals, so that all divergent integrals are the same as those that
appear in vacuum, i.e., at $T=0$ and $\mu_5=0$. This is motivated by a
similar situation in studies of color superconductivity with NJL
models, in that the traditional treatment based on cutoff
regularization leads to a decreasing superconducting gap for high
$\mu$, while the separation of vacuum effects from $\mu$-dependent
divergent integrals leads  to results in agreement with
model-independent predictions~\cite{Farias:2005cr}.  We show  that a
similar effect is at play here, since~$\mu_5$ appears explicitly in
divergent integrals. As such, a decreasing $T_c$ with $\mu_5$ seems to
be a result of improper separation  of medium effects from the vacuum
contributions, thus subject to a dependence on how these  divergent
terms are regularized. This is also similar to the case of magnetized
quark matter,  where unphysical spurious  effects are eliminated by
properly disentangling the magnetic field contributions from
divergent integrals~\cite{mfir,Duarte:2015ppa}. 

Our regularization procedure 
in expressing all divergent  integrals in terms of integrals that appear
in the vacuum is very simple and, once the  divergent vacuum integrals
are fixed to reproduce physical quantities in vacuum, our results
predict an increasing $T_c$ with $\mu_5$. This result is a simple consequence 
of the ability  of writing all divergent integrals in terms of integrals that
appear in the vacuum.  Although  we use a NJL model {\textemdash} see,
e.g. Refs.~\cite{klevansky,buballa} for  reviews and
references{\textemdash} as an explicit example, the procedure applies
equally  well for other effective models for QCD, like the Polyakov--Nambu-Jona-Lasinio (PNJL)
model~\cite{pnjl} that includes the  Polyakov loop contribution.

The remainder of this paper is organized as follows. In
Sec.~\ref{model} we describe the regularization scheme that makes the
vacuum ultraviolet momentum terms independent of the medium effects
and its  implementation in the context of the NJL model at finite
chiral chemical potential and  temperature.  In Sec.~~\ref{numresults}
we contrast the results obtained in the context of this medium
separation scheme with the traditional cutoff one.  Our conclusions
and final remarks are presented in Sec.~\ref{conclusions}.

\section{The NJL model with a chiral imbalance}
\label{model}
  
The NJL model, with a chiral chemical potential included, has the
Lagrangian density given by

\begin{eqnarray}
\mathcal{L} &=& \bar{\psi}  \left(i {\rlap/\partial} - m_c +
\mu_5\gamma^0\gamma^5\right) \psi  \nonumber \\ & +&   G
\left[\left(\bar{\psi}\psi\right)^2    +
  \left(\bar{\psi}i\gamma_5\vec{\tau}\psi\right)^2\right],
\label{lag1}
\end{eqnarray}
where $G$ is the coupling, $m_c$ is the current quark mass ($m_c=0$ in
the quiral limit) and $\psi$ represents a flavor isodoublet,
$N_c$-plet quark  field {\textemdash} a sum over flavors, $N_f=2$, and
color degrees of freedom, $N_c=3$,  is implicit. The mean-field
thermodynamic potential $\Omega(M,T,\mu_5)$ for the model  is a
function of the dynamical quark mass $M \equiv M(T,\mu_5)$, given by the
gap equation  $M = m_c - 2 G \, \langle \bar\psi\psi\rangle$, as  

\begin{eqnarray}
\lefteqn{\Omega(M,T,\mu_5) = \Omega_0(M,\mu_5) } \nonumber\\  &&- 2
N_f N_c T  \sum_{s=\pm 1}  \int \frac{d^3k}{(2\pi)^3} \ln\! \left[ 1 +
  e^{-\omega_s(k)/T} \right],
\label{potential}
\end{eqnarray}
where $\Omega_0$ has no explicit $T$ dependence,

\begin{equation}
\Omega_0(M,\mu_5) \!=\! \frac{(M \!-\! m_c)^2}{4G} \!-\!  N_f N_c
\sum_{s=\pm 1}\int\frac{d^3k}{(2\pi)^3}\omega_s(k),
\label{Omega0}
\end{equation}
and $\omega_s(k)= \sqrt{(|{\bf k}| + s \mu_5)^2+M^2}$  are the
eigenstates of the Dirac operator with helicity $s = \pm 1$.   Note
that while the second term on the right-hand side of
Eq.~(\ref{potential}) is  ultraviolet (UV) finite,  the momentum
integral in $\Omega_0$ is UV divergent and requires a regularization
prescription. $\Omega_0$ depends explicitly on $\mu_5$ and implicitly on $T$,
through its dependence on $M$. To analyze
the gap equation, one will need an integral that is the derivative
with respect to  $M^2$ of the momentum integral in Eq.~(\ref{Omega0});
it can be expressed in  the form

\begin{equation}
\!\!\!\!\frac{\partial}{\partial M^2}\!\left[
  \int\frac{d^3k}{(2\pi)^3}\omega_s(k) \right]\!=\!
\int_{-\infty}^{+\infty} \!\!\frac{dk_4}{2\pi} \! \!
\int\frac{d^3k}{(2\pi)^3} \frac{1}{k_4^2 \!+\! \omega_s^2(k)},
\label{deromega}
\end{equation}
where we have introduced the four-momentum component $k_4$ (in
Euclidean space) for  convenience. In order to  make explicit the
vacuum contribution to the integral, we use three times in  sequence
the identity~\cite{bat1}

\begin{eqnarray}
\frac{1}{k^2_{4} + \omega^2_s(k)}  &=& \frac{1}{k^2_{4}  + k^{2} +
  M^2_{0}}  \nonumber \\ && + \frac{k^2 + M^2_0 -
  \omega^2_s(k)}{\left(k^2_{4} + k^{2} + M^2_{0}\right) \left[k^2_{4}
    \!+\!\omega^2_s(k)\right]},
\label{ident0}
\end{eqnarray}
such that the integrand in Eq.~(\ref{deromega}) can be rewritten
in the form~\cite{Farias:2005cr}

\begin{eqnarray}
\lefteqn{\frac{1}{k^2_{4} + \omega^2_s(k)}  = \frac{1}{k^2_{4}  +k^{2}
    + M^2_{0}} - \frac{A_s(k)}{\left(k^2_{4} + k^{2} +
    M^2_{0}\right)^{2}} } \nonumber \\ && +
\frac{A^2_s(k)}{\left(k_{4}^{2}\! + \! k^{2}\! + \!M^2_{0}
  \right)^{3}} \!-\!  \frac{A^3_s(k)}{\left(k_{4}^{2}\! + \!  k^{2}
  \!+\! M^2_{0} \right)^{3} \left[k^2_{4} \!+\!\omega^2_s(k)\right]},
\label{ident}
\end{eqnarray}
where we have defined $A_s(k) = \mu_{5}^{2} + 2sk \mu_{5} + M^{2} - M^2_{0}$ 
and $M_0$ is the quark mass in the vacuum (i.e., computed at $T=0,
\,\mu_5=0$). Equation~(\ref{ident})  can be verified by direct
algebraic manipulation. Note that, when substituting it back of
Eq.~(\ref{deromega}), the first term on the right-hand  side in
Eq.~(\ref{ident}) leads to a quadratically divergent integral, the two
next terms are proportional to a logarithmically divergent integral,
and the last term leads to a finite integral. It is important to note
that the  divergent integrals are the same as those arising in the
vacuum, as there is  no explicit or implicit dependence on $T$ or
$\mu_5$ in their integrands.  Thus, one can regularize the integrals
as we wish, as, e.g., by a  three-dimensional momentum cutoff
$\Lambda$, and fix $\Lambda$ by fitting a vacuum physical
quantity. The last term,  being finite, can be integrated without any
momentum cutoff, the same way as we do for the second term  
of Eq.~(\ref{potential}), the explicitly temperature dependent term.  

It is at this point where our approach  differs from all previous
calculations: In the traditional approach, the left-hand side of the
identity in Eq.~(\ref{ident}) is used in Eq.~(\ref{deromega}) and a
momentum  cutoff is used to perform the integral with an integrand
that depends explicitly and  implicitly on medium quantities, $\mu_5$
and $M \equiv M(T,\mu_5)$, while when using the  right-hand side of the
identity, Eq.~(\ref{ident}), one obtains divergent integrals that are  
independent from
the medium, i.e., they are dependent on the vacuum quark mass $M_0$
only.  In other words, by using the identity in Eq.~(\ref{ident}),
medium and vacuum dependences can be explicitly disentangled from the
integrands of the  divergent integrals and, therefore, do not get
cutoff by any regulator. In the rest of  this work we refer to this
regularization procedure as ``medium separation scheme'' (MSS),  while
the usual treatment of the divergent integrals as ``traditional
regularization scheme''  (TRS). 

Earlier works that have applied TRS in different effective models of
QCD~\cite{Ruggieri:2011xc,Gatto:2011wc,fukushimaPNJL,huang13,huang14,huang15,chernodub}
have found a critical temperature $T_c$ for chiral symmetry restoration  
that decreases with $\mu_5$. They also find a CEP
on the phase diagram $(\mu_5,T_c)$. Recent lattice
results~\cite{Braguta:2015owi,Braguta:2015zta} obtained instead a
$T_c$ increasing with $\mu_5$ and a transition that is only a
crossover. The idea behind the MSS method is not new~\cite{bat1}, as already mentioned, 
it was used previously in a similar situation that occurs with the NJL in the study
of color superconductivity~\cite{Farias:2005cr}, and it actually resembles~\cite{Sampaio:2002ii} 
the Bogoliubov, Parasiuk, Hepp, Zimmermann renormalization scheme~\cite{Collins:1984xc},
in that the integrand of a divergent amplitude is manipulated to isolate the 
divergence without applying an explicit regulator. 

The dynamical quark mass $M$ is determined
self-consistently by solving the gap  equation derived from
Eq.~(\ref{potential}) which, with the help of Eq.~(\ref{ident}),
becomes

\begin{eqnarray}
\lefteqn{\frac{M \!-\! m_c}{4 N_f N_c \, G M} \!=\! I_{\rm
  quad}\left(\Lambda,M_{0}\right)} 
\nonumber \\ &&  +
\left(2\mu_{5}^{2}\!-\!M^{2}\!+\!M_{0}^{2}\right) I_{\rm
  log}\left(\Lambda,M_{0}\right) \nonumber \\ & -&
\frac{2\mu_{5}^{2}+M^{2}-M_{0}^{2}}{8 \pi^2}
+\frac{M^{2}-2\mu_{5}^{2}}{8\pi^2}\ln\left(\frac{M^2}{M_{0}^{2}}\right)
\nonumber \\ &-& \sum_{s=\pm 1}\int\frac{d^3
  k}{(2\pi)^3}\frac{1}{\omega_s(k)} \frac{1}{e^{\omega_s(k)/T}+1},
\label{Gapsubt}
\end{eqnarray}
where $I_{\rm quad}\left(\Lambda,M_{0}\right)$ and $I_{\rm
  log}\left(\Lambda,M_{0}\right)$ denote the quadratically and
logarithmically UV divergent integrals, respectively,

\begin{eqnarray}
I_{\rm quad}\left(\Lambda,M_{0}\right) =   \int^\Lambda\frac{d^4
  k}{(2\pi)^4}\frac{1}{k^2_4 + k^2 + M_{0}^{2}},
\label{Iquad} 
\end{eqnarray}
and
\begin{eqnarray}
I_{\rm log}\left(\Lambda,M_{0}\right) =   - \frac{\partial}{\partial
  M_0^2} \, I_{\rm quad}\left(\Lambda,M_{0}\right),
\label{Ilog}
\end{eqnarray}
where $\Lambda$ denotes the regularization parameter used in the
divergent integrals. Note that the quark mass dependence of both 
$I_{\rm quad}\left(\Lambda,M_{0}\right)$ and 
$I_{\rm log}\left(\Lambda,M_{0}\right)$ is through the vacuum
quark mass $M_0$.  We reiterate that once a regularization scheme is
chosen, $I_{\rm quad}$ and $I_{\rm log}$ are fixed  by fitting vacuum
properties; for example, $I_{\rm quad}$ and $I_{\rm log}$ can  be
expressed in terms of the quark condensate $\langle {\bar q}
q\rangle$,  the leptonic decay constant $f_\pi$ and the pion mass
$m_\pi$.  Once $G$, $m_c$ and the dynamical quark mass in the vacuum
$M_0$ are  chosen to fit those physical quantities, the integrals are
fixed. 

\section{Numerical Results}
\label{numresults}

\begin{figure}[htb!]
\vspace{0.73cm} \includegraphics[scale=0.34]{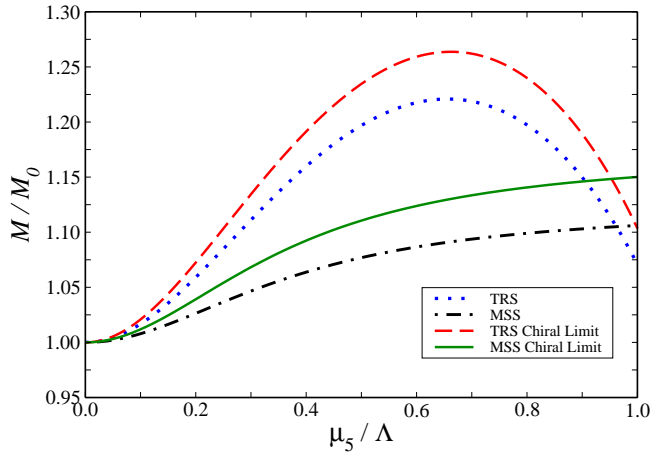}
\caption{The zero temperature quark mass $M$, normalized by its
  respective vacuum value $M_0$, as a function of chiral chemical
  potential  $\mu_5$.  The results obtained using the TRS
  regularization (see text) are given  by the dashed (chiral limit)
  and dotted lines. In the MSS regularization the results  are given
  by the solid (chiral limit, $m_c=0$) and dash-dotted lines. }
\label{fig1}
\end{figure}

We fix the parameters of the model by using as input
$f_{\pi}=92.3$~MeV, $m_{\pi}=0.140$~GeV and
$\langle\bar{q}q\rangle^{\frac{1}{3}}= - 0.250$ GeV, and use a
three-dimensional cutoff to evaluate the vacuum divergent integrals.
A good fit is obtained with $m_c = 5.37$~MeV, $G=4.75$~GeV$^{-2}$ and
$\Lambda = 0.660$~GeV.  The constituent quark mass is found to be
$M_0=0.302$~GeV.

\begin{figure}[htb]
\vspace{0.73cm} \includegraphics[scale=0.34]{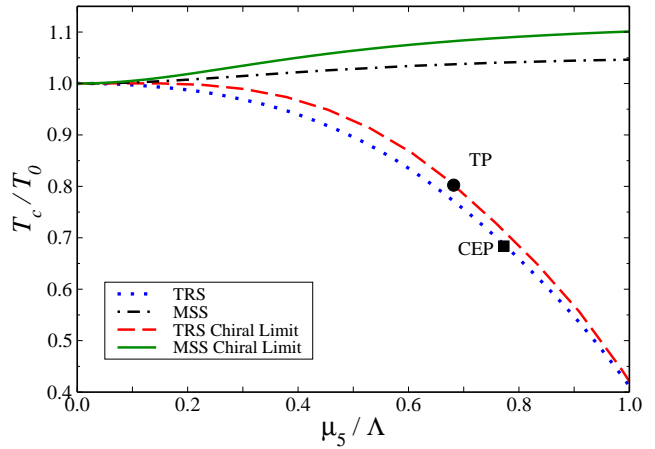}
\caption{The critical temperature $T_{c}$,
 normalized by $T_0 =T_c(\mu_5=0)$), as a function of $\mu_5$.
  The black dot indicates a  tricritical point in the chiral limit,
  while the square indicates the critical end point, both in the TRS
  regularization case (see text for a detailed explanation).}
\label{fig2}
\end{figure}

In {}Fig.~\ref{fig1} we show the results for the dynamical quark mass
$M$ as a  function of $\mu_5$ in the case where $T =0$. The results at
a fixed temperature (below the critical temperature for chiral symmetry
restoration) are qualitatively similar.  We show the results for both
the TRS and MSS regularizations explained above. Here a note of caution
is in order regarding values of $\mu_5$ close to $\Lambda$.
One should keep in mind that the NJL model, being a nonrenormalizable 
model, has an intrinsic energy scale and its predictions of phenomena driven by dynamics 
occurring at energies higher than that scale should be taken with great caution. 
Although the precise limit of validity can be a matter of discussion, as it 
might depend on type of observable or physical process at study, the value for 
that scale is commonly assumed in the literature to be the cutoff~$\Lambda$. In 
view of this and in order to avoid misinterpretations, we have restricted the value 
of $\mu_5$ in {}Fig.~\ref{fig1} to be at most 
$\Lambda$. Note that even though the TRS scheme seems to indicate that the chiral 
chemical potential initially strengthens dynamical chiral symmetry breaking (DCSB), 
the behavior changes at around $\mu_5 \simeq 0.6 \Lambda$, beyond which it starts to 
disfavor DCSB. However, in the MSS scheme, DCSB is always strengthened by the chiral 
chemical potential; with all the required proviso just mentioned, we remark that this 
continues to be true for values of $\mu_5$ larger than $\Lambda$.
Thus, we see that in the TRS regularization, the
tendency of the chiral chemical potential is to weaken the chiral symmetry
breaking beyond $\mu_5 \gtrsim 0.6 \Lambda$, while in the MSS regularization the tendency is
always to strengthen it. This change of behavior, which is directly related on
how the vacuum dependent term on $\mu_5$ is handled, of course
reflects on how the critical temperature changes too. This is
explicitly shown in {}Fig.~\ref{fig2}. 

The values of $T_0$, for the critical ($T_c$) and pseudo-critical
($T_{pc}$) temperatures  for chiral symmetry restoration evaluated at
$\mu_5=0$ used in {}Fig.~\ref{fig2}, are given in Tab.~\ref{table1}. 

\begin{table}[ht]
\caption{\label{pnjl} Values of critical ($T_c$) and pseudo-critical
  ($T_{pc}$) temperatures  for the chiral symmetry restoration at
  $\mu_5=0$}
\begin{center}
\begin{tabular}{ccc}
  \hline
\hspace*{.2cm}  \hspace*{.2cm} & \hspace*{.2cm} $T_c$
(GeV) \hspace*{.2cm}& \hspace*{.2cm} $T_{pc}$ (GeV)  \\ \hline TRS   &
0.165 & 0.177 \\ MSS    & 0.169 & 0.183 \\ \hline
\end{tabular}
\end{center}
\label{table1}
\end{table}

In {}Fig.~\ref{fig2} we show the results for the critical temperature
$T_c$ as a  function of $\mu_5$ for the two forms of treating the
divergent integrals. In the TRS regularization, 
we find a critical end point (CEP) that separates a crossover line from a first-order
transition. In the chiral limit ($m_c=0$) it is instead a tricritical
point (TP), which separates a line of second-order phase transition
from one of   first-order. However, in the MSS regularization both the
TP and the CEP are absent. The  transition is a crossover (note that
in this case $T_c$ in {}Fig.~\ref{fig2} indicates, technically, the
pseudo-critical temperature), while in the chiral limit the transition
is second order throughout. In conformity with the behavior seen for
the dynamical quark mass in {}Fig.~\ref{fig1}, because of the
deleterious effect of the chiral chemical potential on the breaking of
chiral symmetry, $T_c$ decreases in the TRS regularization. But in the
MSS regularization one sees that $T_c$ always increases with
$\mu_5$. This is in qualitative accordance with the recent results from the
lattice~\cite{Braguta:2015owi,Braguta:2015zta} and also with more
sophisticated nonperturbative treatments, e.g., like the ones used in
Refs.~\cite{Wang:2015tia,Xu:2015vna}.  As far the absence of the 
TP (in the chiral limit) or the CEP  in the MSS regularization is
concerned,  this is also seen in the results obtained from
the earlier lattice results~\cite{Yamamoto:2011gk} and also with  the
more recent ones, where no CEP (or TP) has been found. One should, however,
mention here that the lattice results in Refs.~\cite{Braguta:2015owi,Braguta:2015zta}
should be taken with some caution, as they were obtained for a
very large pion mass, $m_\pi = 418$ MeV, while here we used the physical 
value of $m_\pi = 140$ MeV. It
is known that some quantities (for example the behavior of the quark 
condensate as a function of an external magnetic field) may depend heavily on the 
pion mass. So we cannot ruled out the possibility that the nonexistence of
a CEP in those lattice results could be an artifact of the large pion masses
used in those numerical studies.
The increase
of the pseudo-critical and the critical (in the chiral limit)
temperatures are again consistent  with the behavior seen for the
dynamical quark mass in the MSS regularization shown in
{}Fig.~\ref{fig1}. 

{}Finally, as already remarked, being the NJL model an effective model, 
it has an intrinsic scale that 
limits its validity. A natural choice for this scale can be taken for example as 
being the regularization or cutoff scale in the present case, $\Lambda$, and we do expect
that the results should be reliable for values of $\mu_5$ not too above this scale.
We note from the results of both {}Figs.~\ref{fig1} and \ref{fig2}
that the differences between the TRS and MSS regularization schemes are already
significant for values of $\mu_5 \ll \Lambda$. In particular, the differences
between the (pseudo-) critical temperature $T_c$ in the TRS and MSS schemes are 
already apparent for values of $\mu_5$ as low as around $\mu_5 \approx 0.3 \Lambda$, 
where the tendency of growth for $T_c$ is already clear.

\section{Conclusions}
\label{conclusions}

Our results show that a way of conciliating results  for the chiral
critical transition line obtained with NJL models and recent lattice
results, when in the presence of a chiral imbalance, might be closely
connected on how the UV momentum integrals are treated in these
models.  These same results also show that one can eliminate this
discrepancy by a proper separation of medium effects  from the integrand of
the divergent integrals that require regularization. All resulting 
divergent integrals are the same as those
that appear in  the vacuum, i.e., at $T=0$ and $\mu_5=0$. By this
proper separation of medium effects from the divergent vacuum
integrals, we have obtained results for the critical temperature
dependence with the chiral chemical potential that are in qualitative
agreement with physical expectations, in that $\mu_5$ is a catalyst of
DCSB~\cite{kotovmu5cat} and, therefore, an increasing  critical temperature 
as a function of $\mu_5$ should be expected. Moreover, our results are in line
with the arguments of Ref.~\cite{kotovmu5cat} that the ultraviolet cutoff $\Lambda$,
used with a TRS, effectively cuts important degrees of freedom near the Fermi surface 
leading to an incorrect result for the critical temperature as a function of $\mu_5$. 
We also have qualitative agreement with lattice results regarding the absence
of a CEP. Note, however, as we have already remarked, the comparison should be taken 
with caution, given the large pion mass used in those lattice studies. 
Likewise, the position and even (non)existence and of a CEP 
can depend heavily on the pion mass. 
Nevertheless, we must also point out that recent studies~\cite{Wang:2015tia,Xu:2015vna}  based on a 
renormalizable, nonperturbative scheme based on the Dyson-Schwinger equations
of QCD also do not find a critical end point  in the phase diagram 
$(T, \mu_5)${\textemdash}see also discussions in Ref.~\cite{craigmu5pnjl}. 
While definite lattice results with physical pions masses 
are still missing, it is fair to say that there is strong evidence that there is no CEP in 
the phase diagram of quark matter with a chiral imbalance.
In the MSS regularization, we found that the transition is a crossover
in the physical case of $m_0\neq 0$, 
while in the chiral limit, $m_0=0$, it is second-order throughout. 

One additional bonus of properly separating medium effects from
divergent vacuum  momentum integrals, is the fact that once the
parameters of the model are chosen to fit physical quantities in
vacuum, the divergent integrals are fixed and they are not changed 
when studying $T$ and $\mu_5$ effects. This is simply a consequence of making
the UV divergent momentum integrals,  $I_{\rm quad}\left(\Lambda,M_{0}\right)$ 
and  $I_{\rm log}\left(\Lambda,M_{0}\right)$, to depend only on vacuum
quantities. Thus, in the present case where we have chosen a three-dimensional
momentum cutoff $\Lambda$ for the UV divergent integrals, both  $I_{\rm quad}$ and  
$I_{\rm log}$ are fixed once the values of $\Lambda$ and $M_0$ are fitted to the
physical quantities. Even though arguments can be made against such a separation
of vacuum and medium effects in the NJL model, we believe that in some cases, 
such a strategy, in the present case given by the MSS regularization scheme, 
seems to be important for capturing the right physics with the model. Though we have offered arguments 
in favor of the MSS procedure, it is clear that more work is welcome, in particular, 
more work on different regulators is needed.

We believe that this same methodology that we have
employed in this work will also be relevant in any other problem
where this mixing of medium and regularization might be present. Our
results, thus, indicate a way  of improving the predictibility of these
effective models, which are so useful  in our effort to explain one of
the most difficult problems in physics today, i.e., the structure of
the QCD phase diagram.  

\acknowledgments

Work partially financed by Conselho Nacional de Desenvolvimento
Cient\'{\i}fico e  Tecnol\'ogico - CNPq, under the Grant
Nos. 305894/2009-9 (G.K.), 475110/2013-7 (R.L.S.F),  232766/2014-2
(R.L.S.F), 308828/2013-5 (R.L.S.F) and 303377/2013-5 (R.O.R.),
Fun\-da\-\c{c}\~ao de Amparo \`a Pesquisa do Estado de S\~ao Paulo -
FAPESP,  Grant No. 2013/01907-0,  Funda\c{c}\~ao Carlos Chagas Filho
de Amparo \`a  Pesquisa do Estado do Rio de Janeiro (FAPERJ), under
grant No. E - 26 / 201.424/2014 (R.O.R.) and CAPES (D.C.D). R.L.S.F.
acknowledges  the kind hospitality of the Center for Nuclear Research
at Kent State University, where part of this work has been
done. R.L.S.F. is also grateful to Michael Strickland  for insightful
comments and suggestions. We thank A. Y. Kotov for comments and 
discussions. 


\end{document}